\documentclass[envcountsect]{llncs}

\usepackage[polutonikogreek,english]{babel}

\def\calF{\mathcal{F}}
\def\calG{\mathcal{G}}

\def\calP{\mathcal{P}}
\def\calS{\mathcal{S}}

\def\NLC{\mathcal{NLC}}
\def\StS{\calS t\calS}
\def\GP{\calG\calP}
\def\GS{\calG\calS}

\def\form#1{\mathbf{#1}}
\def\DA{\form{DA}}

\def\val{\textsf{val}}
\def\parse{\textsf{parse}}

\def\FO{\textsf{FO}}
\def\MS{\textsf{MS}}
\def\MSO{\textsf{MSO}}
\def\CMSO{\textsf{CMSO}}
\def\PTL{\textsf{PTL}}
\def\MOD{\textsf{MOD}}
\def\inc{\textsf{inc}}
\def\edge{\textsf{E}}

\def\inv{^{-1}}
\let\phi\varphi

\let\N\bbbn

\begin{document}

\title{Algebraic Recognizability of Languages\thanks{This paper was
prepared in part while the author was an invited Professor at the
University of Nebraska-Lincoln. He also acknowledges partial support
from project \textsc{versydis} (\textit{ACI S\'ecurit\'e Informatique},
Minist\`ere de la Recherche).}}

\author{Pascal Weil}

\tocauthor{Pascal Weil (LaBRI - CNRS)}

\institute{LaBRI, CNRS and Universit\'e Bordeaux-1,
\email{pascal.weil@labri.fr}}

\maketitle

\begin{abstract}
    Recognizable languages of finite \textit{words} are part of every
    computer science cursus, and they are routinely described as a
    cornerstone for applications and for theory. We would like to briefly
    explore why that is, and how this word-related notion  extends
    to more complex models, such as those developed for modeling
    distributed or timed behaviors.
\end{abstract}

\begin{otherlanguage}{polutonikogreek}
    \noindent >En >arq~h| >~hn <o l'ogos\dots
\end{otherlanguage}

\noindent \textsl{In the beginning was the Word\dots}

\bigskip

Recognizable languages of finite \textit{words} are part of every
computer science cursus, and they are routinely described as a
cornerstone for applications and for theory. We would like to briefly
explore why that is, and how this word-related notion extends to more
complex models, such as those developed for modeling distributed or
timed behaviors.

The notion of recognizable languages is a familiar one, associated with
classical theorems by Kleene, Myhill, Nerode, Elgot, B\"uchi,
Sch\"utzenberger, etc. It can be approached from several angles:
recognizability by automata, recognizability by finite monoids or
finite-index congruences, rational expressions, monadic second order
definability. These concepts are expressively equivalent, and this
leads to a great many fundamental algorithms in the fields of
compilation, text processing, software engineering, etc\dots\ Moreover,
it surely indicates that the class of recognizable languages is
central. These equivalence results use the specific structure of words
(finite chains, labeled by the letters of the alphabet), and the monoid
structure of the set of all words.

Since the beginnings of language theory, there has been an interest for
other models than words -- especially for the purpose of modeling
distributed or timed computation (trees, traces, pomsets, graphs, timed
words, etc) --, and for extending to these models the tools that were
developped for words. For many models, some of these tools may not be
defined, and those who are defined, may not coincide.

In this paper, we concentrate on the algebraic notion of
recognizability: that which, for finite words, exploits the monoid
structure of the set of words, and relies on the consideration of
monoid morphisms into finite monoids, or equivalently, of finite-index
monoid congruences. Our aim is to examine why this particular approach
is fruitful in the finite word case, and how it has been, or can be
adapted to other models.

In Sect. \ref{sec words}, we explore the specific benefits of
using algebraic recognizability for the study of word languages. It
opens the door to a very fine classification of recognizable
languages, which uses the resources of the structural theory of finite
monoids. This classification of recognizable languages is not only
mathematically elegant, it also allows the characterization and the
decision of membership in otherwise significant classes of languages.

The emblematic example of such a class is that of star-free languages,
which are exactly the first-order definable languages, those that can
be defined by a formula in propositional temporal logic, those that are
recognized by a counter-free automaton, and those whose syntactic
monoid is aperiodic (Sch\"utzenberger, McNaughton, Pappert, Kamp). Only
the latter two characterizations lead to a decision algorithm, and the
algebraic approach makes this algorithm the clearest.

The example of star-free languages is however not the only one; for
example, the various notions of locally testable languages are also
characterized, and ultimately efficiently decided, by algebraic
properties (Simon, McNaughton, Ladner). The power of finite monoid
theory leads in fact to an extremely fine classification (Eilenberg),
where for instance, natural hierarchies within first-order or temporal
logic can be characterized as well (Brzozowski, Knast, Thomas,
Th\'erien, Wilke).

Already in the 1960s, fundamental results appeared on notions of
auto\-mata for trees and for infinite words, linking them with logical
definability and rational expressions (B\"uchi, Doner, Mezei, Thatcher,
Wright). An algebraic approach to automata-recognizable languages of
infinite word was introduced in the early 1990s, and as in the case of
tree languages, it requires introducing an algebraic framework
different from monoid theory (Perrin, Pin, Wilke), see Sect.~\ref{sec
infinite words}.

In fact, very early on, Elgot and Mezei extended the notion of
recognizability to subsets of arbitrary (abstract) algebras. But the
notion of logical definability for these subsets strongly depends on
the combinatorial (relational) structure of the objects chosen to
represent the elements of the abstract algebra under consideration. In
many situations, the problem is posed in the other direction: we know
which models we want to consider (they are posets, or graphs, or
traces, or timed words, as arise from, say, the consideration of
distributed or timed computation) and we need to identify an algebraic
structure on the set of these objects, for which logical definability
and algebraic recognizability will be best related. One key objective
there, is to be able to decide logical specifications. Note that models
of automata, while highly desirable, are not known to exist in all the
interesting cases, and especially not for graphs or posets. In
contrast, the algebraic and the logical points of view are universal.

A class of relational structures being fixed, Courcelle gave very loose
conditions on an algebraic structure on the set of these structures, 
which guarantee that counting monadic second order definability
implies recognizability. The converse is known to hold in a number of
significant cases, but not in general.

In Sect. \ref{successful examples}, we discuss some of the relational
structures that have been studied in the literature, and the
\textit{definability vs. recognizability} results that are known for
them: trees, infinite words, traces, series-parallel pomsets, message
sequence charts and layered diagrams, graphs, etc. The main obstacle to
the equivalence between definability and recognizability is the fact
that the algebras we consider may not be finitely generated. In
contrast, a good number of situations have been identified where this
equivalence holds, and each time, a finiteness (or boundedness)
condition is satisfied. We will try to systematically point out these
finiteness conditions, and we will identify some important questions
that are still pending.

When all is said and done, the central question is that of the
specification and the analysis of infinite sets by finite means (finite
and finitely generated algebras, finite automata, logical formulas,
etc.), arguably the fundamental challenge of theoretical computer
science. This paper presents a personal view on the relevance of
algebraic recognizability for this purpose, beyond its original scope
of application (languages of finite words), and an introduction to some
of the literature and results that illustrate this view. I do not claim
however that it constitutes a comprehensive survey of the said
literature and results (in particular, I chose to systematically refer
to books and survey papers when available), and I apologize in advance
for any omission!\dots

\section{The Finite Word Case}\label{sec words}

In the beginning were the (finite) words, and loosely following the
Biblical analogy, one could say that the spirits of Kleene, B\"uchi and
Sch\"utzenberger flew over the abyss, organising it from chaos to
beauty.

Throughout this paper, $A$ will denote an alphabet, i.e. a finite,
non-empty set. We denote by $A^*$ the set of all finite words on
alphabet $A$.

\subsection{The Classical Equivalence Results}

We are all familiar with the notion of regular languages, but there are
in fact several competing notions, that turn out to be equivalent for
finite words. Each is interesting in its own right, as it reveals a
fruitful point of view, syntactic or semantic, denotational or
operational. The results of this section can be found in many books,
and in particular in those of Eilenberg \cite{Eilenberg}, Pin
\cite{Pin86,Pin1996,Pin1997}, Straubing \cite{Straubing}, Sipser
\cite{Sipser} and Sakarovitch \cite{Sakarovitch}.

\subsubsection{Recognizability by Automata.}
One can first consider languages \textit{recognized by finite state
automata}, whether deterministic or non-deterministic. Every language
recognized by a finite state automaton admits a unique \textit{minimal
deterministic automaton}, which is effectively computable.

The notion of a deterministic automaton can also be expressed in terms
of a finite-index semi-congruence, and in terms of an action of the
free monoid on a finite set.

\subsubsection{Algebraic Recognizability.}
One can also consider languages \textit{recognized by a finite monoid}.
This exploits the monoid structure of $A^*$, the set of all words on
alphabet $A$: if $M$ is a monoid and $\phi\colon A^* \rightarrow M$ is
a morphism of monoids, we say that $L\subseteq A^*$ is recognized by
$\phi$ (or by $M$) if $L = P\phi\inv$ for some $P\subseteq M$, or
equivalently, if $L = L\phi\phi\inv$. Here too, for every language $L$
recognized by a finite monoid, there exists a least finite monoid
recognizing $L$, called the \textit{syntactic monoid} of $L$.

\subsubsection{Rational Expressions.}
\textit{Rational expressions} describe languages using the letters of
the alphabet, the constant $\emptyset$, and the so-called
\textit{rational operations} of union, concatenation and Kleene star
(if $L$ is a language, $L^*$ is the submonoid of $A^*$ generated by
$L$).

It should be noted that for a given rational language, there is no
notion of a unique minimal rational expression describing it.

\subsubsection{Logical Definability.}
\textit{B\"uchi's sequential calculus} exploits the combinatorial
structure of words, as $A$-labeled, linearly ordered finite sets: in
this logical formalism, individual variables are interpreted to be
positions in a word, and the predicates are $i<j$ (to say that position
$i$ is to the left of position $j$) and $R_{a}i$ (to say that position
$i$ is labeled by letter $a\in A$). \textit{First order formulas}
($\FO$) use only individual variables, whereas \textit{monadic second
order formulas} ($\MSO$) also use second order variables, interpreted to
be sets of positions. To a formula $\phi$ in this language, one
associates the set $L(\phi)$ of all words which satisfy $\phi$, that is
$L(\phi)$ is the language of the finite models of $\phi$.

\subsubsection{The Kleene-Nerode-Myhill-B\"uchi Theorem.}
Theorems by Kleene, Ner\-ode, Myhill and B\"uchi show that the notions
rapidly described above coincide.

\begin{theorem}\label{KNB}
    Let $L \subseteq A^*$. Then $L$ is recognized by a finite state
    automaton, if and only if $L$ is recognized by a finite monoid, if
    and only if the syntactic monoid of $L$ is finite, if and only if
    $L$ is described by a rational expression, if and only if $L$ is
    defined by an $\MSO$-formula of B\"uchi's calculus.
    
    Moreover, there are algorithms to pass from one of these
    specification formalisms to each other.
\end{theorem}

It is interesting to note that the many closure properties of the class
of recognizable languages are easily established in an appropriate
choice of one of these equivalent formalisms. For instance, closure
under Boolean operations easily follows from the definition of
algebraic recognizability, as does closure under inverse morphism
(inverse rewriting). On the other hand, closure under concatenation,
star and direct morphism is a triviality for languages described by
rational expressions.

None of the equivalences in Theorem~\ref{KNB} is very difficult, but
their proofs really use the different points of view on words and
languages. If we compare these results with the situation that prevails
for other models than words, it is in fact a very exceptional situation
to have these notions be so nicely defined and be equivalent.

\subsection{Classification of Recognizable Languages}\label{sec classification}

With each recognizable language $L$, we can associate a computable
canonical finite object -- in fact two closely related such objects:
the minimal automaton of $L$, and its syntactic monoid. The connection
between them is tight: the syntactic monoid of $L$ is exactly the
monoid of transitions of its minimal automaton.

This paves the way for a fine classification of recognizable languages
(see \cite{Pin86,Pin1997}). Not surprisingly, it is the syntactic
monoid, with its natural algebraic structure, which offers the
strongest classification tool. In this section, we give a few instances
of this classification: some are well-known and open up important
applications (star-free languages, locally testable languages), some
are more specific, and demonstrate the degree of refinement allowed by
this method.

\subsubsection{Star-Free Languages.}
The most illuminating example of a significant subclass of recognizable
languages is given by the \textit{star-free languages}. These are the
languages which can be described by star-free expressions,
\textit{i.e.}, using the letters of the alphabets, the constants
$\emptyset$ and $1$ (the empty word), the Boolean operations and
concatenation (but no star).


The characterization of star-free languages requires the following
definitions: a deterministic automaton is said to be
\textit{counter-free} if whenever a non-trivial power $u^n$ ($u\ne 1$,
$n\ne 0$) labels a loop at some state $q$, then the word $u$ also
labels a loop at the same state. A finite monoid $M$ is said to be
\textit{aperiodic} if it contains no non-trivial group, if and only if
for each $x\in M$, $x^n = x^{n+1}$ for all large enough $n$. Finally,
$\PTL$ (\textit{propositional temporal logic}) is a modal logic,
interpreted on positions in words, with modalities \textsf{next},
\textsf{eventually} and \textsf{until}.

The following statement combines results by Sch\"utzenberger,
McNaughton, Pappert and Kamp, see
\cite{Eilenberg,Kamp,Pin86,Straubing}.

\begin{theorem}\label{SMcNK}
    Let $L \subseteq A^*$. Then $L$ is star-free, if and only if $L$ is
    recognized by a finite aperiodic monoid, if and only if the
    syntactic monoid of $L$ is finite and aperiodic, if and only if $L$
    is recognized by a counter-free automaton, if and only if $L$ is
    defined by an $\FO$-formula of B\"uchi's calculus, if and only if
    $L$ is defined by a $\PTL$-formula.
    
    Moreover, there are algorithms to pass from one of these
    specification formalisms to each other.
\end{theorem}

Thus, the class of star-free languages, with its natural definition in
terms of generalized rational expressions, ends up having natural
characterizations in terms of all the formalisms used in
Theorem~\ref{KNB} -- to which we can add $\PTL$, a logical formalism
considered to be very useful to specify the behavior of complex
systems.

The historically first side of this result is the algebraic one, which
links star-free languages and aperiodic monoids. It is of particular
interest for two reasons. First because it offers an algorithm to decide
whether a language is star-free; and second, because it shows that the
algebraic structure of the syntactic monoid of a recognizable language
(not just its finiteness) reflects the combinatorial properties of that
language. This gave the first hint of Eilenberg's theorem, discussed
further in this section.

\subsubsection{Variants of $\FO$-Definability.}
Several refinements and generalizations further reinforce the significance of
Theorem~\ref{SMcNK}.

Consider the extension $\FO{+}\MOD$ of $\FO$, where we also allow
\textit{modulo quantification} of the form $\exists^{\bmod q}x\
\phi(x)$ ($q\ge 1$). Such a quantification is interpreted to mean that
the set of values $x$ for which $\phi(x)$ holds, has cardinality a
multiple of $q$. Straubing, Th\'erien and Thomas showed that a language
is $\FO{+}\MOD$-definable if and only if the subgroups of its syntactic
monoid are solvable \cite{Straubing}.

Considering now subclasses of $\FO$, it turns out that every star-free
language can be defined by a $\FO$-formula using only 3 variables. Let
$\FO_{2}$ be the class of star-free languages defined by $\FO$-formulas
with only 2 variables. Let also $\DA$ be the class of finite monoids in
which every regular element is idempotent (if $xyx = x$ for some $y$,
then $x^2 = x$). Then a combination of results of Etessami, Pin,
Sch\"utzenberger, Th\'erien, Vardi, Weil, Wilke
\cite{Sch76,PinW1997,EVW2002,ThWi1998} shows that a language $L$ is in
$\FO_{2}$, if and only if it is defined by a $\Sigma_{2}$- and by a
$\Pi_{2}$-formula, if and only if its syntactic monoid is in $\DA$, if
and only if $L$ is defined by a $\PTL$ formula which does not use the
\textsf{until} modality, if and only if $L$ can be obtained from the
letters using only disjoint unions and unambiguous products (and the
constants $\emptyset$ and $A^*$).

It is well-known that every $\FO$-formula is equivalent to one in
prenex normal form (consisting of a sequence of quantifications,
followed by a quantifier-free formula). This gives rise to the
classical \textit{quantifier-alternation hierarchy}, based on counting
the number of alternated blocks of existential and universal
quantifiers. Another natural hierarchy, seen from the point of view of
star-free expressions, defines its $n{+}1$-st level as the Boolean
closure of products of level $n$ languages (and level 0 consists of
$\emptyset$ and $A^*$). This is the so-called \textit{dot-depth
hierarchy}. Thomas showed that these two hierarchies coincide, that is,
a language can be defined by an $\FO$-formula in prenex normal form
with $n$ alternating blocks of quantifiers, if and only if it is in the
$n$-th level of the dot-depth hierarchy \cite{Thomas82}. Decidable
algebraic characterizations were given for level 1 of these
hierarchies, but the decidability of level 2 and the further levels is
still an open question. It was however showed (Brzozowski, Knast,
Simon, Straubing, see \cite{Pin1996}) that the hierarchy is infinite (if
$|A|\ge 2$), and that each level is characterized by an algebraic
property, in the following sense: if two languages have the same
syntactic monoid and one is at level $n$, then so is the other one.

There is also a natural hierarchy on $\PTL$-formulas, based on the
number of nested usage of the $\textsf{until}$ modality. Th\'erien and
Wilke showed that the levels of this infinite hierarchy are
characterized by the algebraic properties of the syntactic monoid, and
that each is decidable \cite{ThWi2002}.

\subsubsection{Communication complexity.}

The \textit{communication complexity} of a language $L$ is a measure of
the amount of communication that is necessary for two partners, each
holding part of a word, to determine whether the word lies in $L$, see
\cite{KushNisan}. Tesson and Th\'erien showed that the communication
complexity of a recognizable language is entirely determined by its
syntactic monoid, and that it can be computed on this basis
\cite{TeTh2003}.

\subsubsection{Piecewise and Locally Testable Languages.}

A word $v$ is a \textit{subword} of a word $u$ if $v = a_{1}\cdots
a_{n}$ and $u = u_{0}a_{1}u_{1}\cdots a_{n}u_{n}$ for some
$u_{0},\ldots,u_{n}\in A^*$. It is a \textit{factor} of $u$ if $u =
xvy$ for some $x,y\in A^*$.

A language $L$ is said to be $n$-piecewise testable if whenever $u$ and
$v$ have the same subwords of length at most $n$ and $u\in L$, then
$v\in L$. The language $L$ is \textit{piecewise testable} if it is
$n$-piecewise testable for some $n$.

A language $L$ is said to be $n$-locally testable if whenever $u$ and
$v$ have the same factors of length at most $n$ and the same prefix and
suffix of length $n-1$, and $u\in L$, then $v\in L$. The language $L$
is \textit{locally testable} if it is $n$-locally testable for some
$n$. Locally testable languages are widely used in the fields of
learning and pattern matching, whereas piecewise testable languages
form the first level of the dot-depth hierarchy.

Results of Simon and McNaughton, Ladner (see
\cite{Eilenberg,Pin86,Pin1997}) show that both these properties are
characterized by algebraic properties of syntactic mon\-oids. More
precisely, a language $L$ is piecewise testable if and only if every
principal two-sided ideal of its syntactic monoid $S(L)$, admits a
single generator. The language $L$ is locally testable if and only if
$S(L)$ is aperiodic and $eS(L)e$ is an idempotent commutative monoid,
for each idempotent $e\ne 1$ in $S(L)$.

\subsubsection{Varieties of Languages.}
Many more examples can be found in the literature, where natural
algebraic properties of finite monoids match natural combinatorial or
logical properties of languages (see for instance
\cite{Almeida94,Eilenberg,Pin86,Pin1997}). The scope of this matching
is described in Eilenberg's variety theorem; the latter identifies the
closure properties on classes of recognizable languages and on classes
of finite monoids, that characterize the classes that can occur in this
correspondence. These classes are called, respectively,
\textit{varieties of recognizable languages} and
\textit{pseudovarieties of finite monoids}.

\subsubsection{Decision Procedures.}

The varieties of recognizable languages thus identified by algebraic
means are all the more interesting if they are decidable. Since the
syntactic monoid of a recognizable language is computable, this reduces
to deciding the membership of a finite monoid in certain
pseudovarieties. In fact, in the examples surveyed above, this is the
only path known to a decision algorithm.

Let us now assume that we are considering a decidable pseudovariety of
monoids (and hence a decidable variety of languages). The syntactic
monoid of a language $L$, which is the transition monoid of the minimal
automaton of $L$, may have a size exponential in the number of states
of that automaton. Thus deciding whether a recognizable language given
by a deterministic finite state automaton lies in a given variety,
seems to require exponential time and space.

In view of the connection between syntactic monoid and minimal
automaton, it is possible to translate the relevant algebraic property
of finite monoids to a property of automata, and to check this property
on the minimal automaton. This possibility is explicitly stated in
Theorem~\ref{SMcNK}, but it is also the underlying reason for the
decision procedures concerning piecewise testable (Stern
\cite{Stern85}) and locally testable languages (Kim, McNaughton,
McCloskey \cite{KMM91}).

In many important situations, this leads to polynomial time membership
algorithms: piecewise and locally testable languages, $\FO_{2}$,
certain varieties related to the dot-depth hierarchy \cite{PinW1997},
etc. One major exception though, is the class of star-free languages,
for which the membership problem is PSPACE-complete (Cho, Huynh
\cite{ChoHuynh}). In other words, given a deterministic automaton,
there is no fundamentally better algorithm to decide whether the
corresponding language is star-free, than to verify whether the
syntactic monoid is aperiodic.

It must be stressed that even in the cases where we have polynomial
membership algorithms, these algorithms are a translation to automata
of algebraic properties of the syntactic monoid, they were not
discovered until after the corresponding pseudovariety of monoids was
identified, and their natural justification is \textit{via}
monoid-theoretic considerations.

\subsection{Recognizable and Context-Free Languages}\label{equational
words}

Recognizable languages form but the lowest level of the Chomsky
hierarchy, where the next level consists of the context-free languages.
\textit{Context-free languages} are defined by context-free grammars,
which can be viewed, with a more algebraic mindset, as finite systems
of polynomial equations of the form $x_{i} = \sum p(\vec x)$ ($1\le
i\le n$) where $\vec x = (x_{1},\ldots,x_{n})$ is the vector of
variables, the summations are finite and each $p$ is a word over the
letters of $A$ and the variables (see \cite{Berstel}). A solution of
such a system is a vector of languages $\vec L = (L_{1},\ldots,
L_{n})$, and the context-free languages arise as the components of
maximal solutions of such systems. Accordingly, context-free languages
are also called \textit{equational}, or \textit{algebraic}.

Recognizable languages are components of maximal solutions of certain
simpler systems, where each $p(\vec x)$ is a word of the form $x_{j}u$,
$1\le j\le n$, $u\in A^*$ (right-linear equation). In particular, not
all context-free languages are recognizable. The class of context-free
languages is not closed under intersection, but it is closed under
intersection with recognizable languages.

\section{Almost as Established: the Finite Tree Case}\label{sec trees}

\textit{Tree languages} were considered in the early 1960s, see
\cite{GecsegSteinby}. Here we use a \textit{ranked alphabet}, that is,
a set $\Sigma$, equipped with an arity function $\sigma\colon
\Sigma\rightarrow\N$. A $\Sigma$-term is defined recursively as
follows: every letter of arity 0 (a constant) is a $\Sigma$-term, and
if $a\in \Sigma$ has arity $n$ and $t_{1},\ldots,t_{n}$ are
$\Sigma$-terms, then $a(t_{1},\ldots,t_{n})$ is a $\Sigma$-term. Terms
are naturally (and unequivocally) represented by $\Sigma$-labeled
trees, where an $a$-labeled node has $\sigma(a)$ linearly ordered
children. We let $T_{\Sigma}$ be the set of all $\Sigma$-terms.

Thatcher and Wright introduced a model of automata for $\Sigma$-labeled
trees, the so-called \textit{bottom-up automata}
\cite{GecsegSteinby,ThatcherWright}. To describe their expressiveness,
they used the natural algebraic structure on the set of $\Sigma$-terms:
each element $a\in \Sigma$ is an operation, of arity $\sigma(a)$, and
no relation is assumed to hold between these operations. Now, let a
$\Sigma$-algebra be any set $S$ equipped with a $\sigma(a)$-ary
operation $a^S$ for each $a\in \Sigma$. If we use this algebraic
notion, we can define recognizable and equational sets of
$\Sigma$-terms: a subset $L\subseteq T_{\Sigma}$ is said to be
recognizable if there exists a morphism (of $\Sigma$-algebras) $\phi$
from $T_{\Sigma}$ to a finite $\Sigma$-algebra, such that $L =
L\phi\phi\inv$; and $L$ is equational if it is a component of a vector
of maximal solutions of a system of polynomial equations. These systems
are defined as in Sect.~\ref{equational words}, except that the
parameters $p$ are now taken to be terms rather than words. Note that
again, given $L\subseteq T_{\Sigma}$, there exists a unique least
$\Sigma$-algebra recognizing it, called the \textit{syntactic
$\Sigma$-algebra} of $L$.

Thatcher and Wright also described subsets of $T_{\Sigma}$ by
\textit{generalized rational expressions}, involving the letters,
unions, the $\Sigma$-operations, and an appropriate notion of
iteration.

Finally, Doner considered a logical formalism to be applied to the
trees representing $\Sigma$-terms \cite{Doner}: the individual variables are
interpreted as nodes in a finite tree and the predicates are
interpreted to express the labeling function and the parent-child
relation.

Results of Doner, Thatcher and Wright
\cite{Doner,GecsegSteinby,ThatcherWright} prove the following
statement.

\begin{theorem}\label{TWD}
    Let $L\subseteq T_\Sigma$. Then $L$ is recognized by a bottom-up
    automaton, if and only if $L$ is recognized by a finite
    $\Sigma$-algebra, if and only if the syntactic $\Sigma$-algebra of $L$ is
    finite, if and only if $L$ is described by a generalized rational
    expression, if and only if $L$ is defined by an $\MSO$ formula, if
    and only if $L$ is equational.
    
    Moreover, there are algorithms to pass from one of these
    specification formalisms to each other.
\end{theorem}

Note that the particularity of this setting is that equational sets are
recognizable. Another important remark is that deterministic bottom-up
automata are really $\Sigma$-algebras, so the notions of
automata-theoretic and algebraic recognizability are not really
distinct. This last point makes Theorem~\ref{TWD} a little less
satisfying than its word counterpart. Another (subjective) cause of
dissatisfaction is that the generalized rational expression are rather
awfully complex. Finally, this result has not made it easy to classify
term languages in the spirit of Sect.~\ref{sec classification}. This is
maybe due to a less long history of investigating the structural
properties of finite $\Sigma$-algebras. Some interesting related
results on binary trees, expressed in terms of certain context-free
languages of words were proved by Beaudry, Lemieux, Th\'erien
\cite{BeaudryITA98,BeaudryTCS98,BLT97,BLT2001}. Nevertheless, it is
fair to say that no structural theory of $\Sigma$-algebras clearly
emerges.

An open question which may serve as a benchmark in this direction is
the following: given a recognizable tree language, can one decide
whether it is $\FO$-definable?

\section{The General Notion of Recognizability}

Adapting the discussion in Sect.~\ref{sec trees}, one can easily define
\textit{recognizable} and \textit{equational} subsets in any algebra.
Recognizable sets are defined in terms of morphisms into finite
algebras of the same type (or in terms of finite index congruences),
and equational sets in terms of systems of polynomial equations (Mezei,
Wright \cite{MezeiWright}, Courcelle \cite{BC1996}). With those definitions, recognizable sets
form a Boolean algebra, equational sets are closed under union,
recognizable sets are always equational, finite sets are equational
(even though they may fail to be recognizable), products of equational
sets (using the operations in the algebra under consideration) are
equational, but the analogous statement for recognizable sets is not
always true, and the intersection of a recognizable and an equational
set is equational. Finally, if $\phi\colon S \rightarrow T$ is a
morphism between algebras of the same type, then $L\phi\inv$ is
recognizable if $L\subseteq T$ is recognizable, and $L\phi$ is
equational if $L\subseteq S$ is equational.

\subsection{Choosing an Algebraic Structure}

As discussed above, if the sets we consider are naturally contained in
an algebra, the notion of recognizability is straightforward. Sometimes
however (frequently maybe), we want to discuss sets of relational
structures, and we then design an algebraic signature to combine these
structures.

For instance, it is one such abstract construction that has us see
trees as terms. Consider even finite words: the interest of the model
maybe lies simply in the notion of a totally ordered $A$-labeled finite
set. We chose to view the set of words as a monoid under concatenation,
and this gave rise to the notion of algebraically recognizable
languages discussed in Sect.~\ref{sec words}. We could also consider
the following algebraic structure on the set of words: each letter
$a\in A$ defines a unary operation $u\mapsto ua$. Then the set of all
finite words is the algebra generated by $A$ and the constant $1$ (this
amounts to considering the set of words as the algebra of
$(A{\cup}\{1\})$-terms). One can verify that the notion of recognizable
language is not modified -- another sign of the robustness of the model
of words. In fact, finite $(A{\cup}\{1\})$-algebras are naturally
identified with deterministic finite state automata, and the
equivalence between this notion of algebraic recognizability and the
monoid-based one is a rephrasing of Kleene's theorem.

Relational structures, in the sense of this paper, are sets equipped 
with relations from a given relational signature. For instance, as
mentioned earlier, words are $A$-labeled totally ordered sets: the
relational signature consists of the binary order relation and one
labeling unary relation $R_{a}$ for each letter $a\in A$. In trees, the
relations are the labeling relations and the parent-child binary
relation (or the predecessor relation, or the parent and the sibling
relations, etc, -- these choices are equivalent when it comes to
expressing properties in monadic second order logic, see
Sect.~\ref{def rec}).

The choice of an algebraic structure can be guided by the natural
constructions generating the finite relational structures under
consideration (concatenation of words; construction of terms;
construction of a word letter by letter), but there is really nothing
canonical or unique about the algebraic structure.

Suppose for instance that we consider very few operations: then there
are many more recognizable set, maybe to the extent that every set is
recognizable (for exemple, consider the set of words with no operations
at all: every finite partition, say $L$ and $L^c$, is a finite index
congruence). If on the other hand we have too many operations, then
there will be less recognizable sets. For instance, let $sh$ (for
\textit{shift}) be the unary operations on words that fixes $1$, and
maps $ua$ to $au$ ($a\in A$, $u\in A^*$). One can verify that the set
$a^*b$ is not recognizable for the algebra whose signature consists of
the concatenation product and the shift operation. Another extreme
example is given by $\N$, equipped with the constant $0$ and the
unary predecessor and successor operations: then the only
recognizable sets are $\emptyset$ and $\N$.

It may also happen that adding certain operations does not change the
class of recognizable sets. For instance, adding the mirror operation
(defined inductively by $\tilde 1 = 1$ and $\widetilde{ua} = a\tilde
u$) to the concatenation product, does not alter the notion of
recognizability. See also Sect.~\ref{sec graph}.

In Sect.~\ref{successful examples}, we discuss a number of relational
structures for which very interesting notions of recognizability have
emerged in the literature.

\subsection{Multi-Sorted Algebras, Ordered Algebras,
etc}\label{multisorted}

Sometimes, algebras are too constrained: the domain and the range of
certain natural operations may consist of certain kinds of elements
only. This is taken care of by the definition of \textit{multi-sorted
algebras}, see \cite{BC1996,DeWi}.

A typical example is provided by the study of languages of infinite
words (more details are given in Sect.~\ref{sec infinite words}). It
turns out that the best algebraic framework consists of considering
simultaneously the finite and infinite words. One relevant operation is
the concatenation product: between two finite words, it is the usual,
fundamental operation, yielding a finite word; the product $uv$ where
$u$ is a finite word and $v$ is infinite, is an infinite word; and
while it is possible to define the product of two infinite words, the
outcome of such a product carries no significant information ($uv=u$)
and the operation can be discarded. So we find that we need to consider
two \textit{sorts} of elements, \textsf{finite} and \textsf{infinite},
and two binary product operations, of type
$\textsf{finite}\times\textsf{finite}\rightarrow\textsf{finite}$ and
$\textsf{finite}\times\textsf{infinite}\rightarrow\textsf{infinite}$.
We also need to consider the $\omega$-power, a unary operation of type
$\textsf{finite}\rightarrow\textsf{infinite}$ (since it turns a finite
word into an infinite one).

Another example is discussed in Sect.~\ref{sec graph}, where algebras
with infinitely many sorts are considered. We do not want to give here
a detailed discussion of congruences in multi-sorted algebra, only
pointing out that such congruences can only identify elements of the
same sort. If there are finitely many sorts, recognizability is defined
by considering morphisms into finite algebras, or finite-index
congruences. If the algebraic signature under consideration has
infinitely many sorts, non-trivial algebras are usually not finite, and
we consider \textit{locally finite} algebras (in which each sort has a
finite number of elements) and \textit{locally finite index}
congruences (with a finite number of classes in each sort).

In the mid-1990s, Pin introduced the usage of \textit{ordered
semigroups} to refine the classification of recognizable languages
\cite{Pin1997}. The same idea can as naturally be used in any algebra
(and has been for instance in \cite{LW-TCS}), but we will keep it
outside the discussion in this paper to avoid increased complexity.

\subsection{Definability vs. Recognizability}\label{def rec}

Based on the examples of words and trees, the natural language for
logical definability of recognizable sets would seem to be $\MSO$,
\textit{monadic second order logic}. It is actually more natural to use
$\CMSO$, \textit{counting monadic second order logic}. $\CMSO$
\cite{BCI,BCHdbk} is monadic second order logic, enriched with the
modulo quantifiers $\exists^{\bmod q}x$ introduced in Sect.~\ref{sec
words}. In the case of words, $\CMSO$ is equivalent to $\MSO$. In fact,
this holds for any relational structure that comes equipped with a
linear order, or for which a linear order can be defined by a
$\MSO$-formula (\textit{e.g.} $A$-labeled trees as in Sect.~\ref{sec
trees}, or traces as in Sect.~\ref{sec poset}), but it is not true in
general.

For instance, when discussing multisets (subsets of $A$ with
multiplicity), we can view them as $A$-labeled finite discrete graphs
(graphs without edges). Then, $\MSO$ can only define finite and
cofinite sets, and it is strictly weaker than $\CMSO$. Note that,
algebraically, the multisets on $A$ under union, form the free
commutative monoid on $A$. The same monoid can be interpreted in terms
of traces (with a commutative alphabet), its elements are then viewed
as certain directed acyclic graphs with one connected component per
letter (see Sect.~\ref{sec poset} on traces), and $\MSO$ is equivalent
to $\CMSO$ in this context. The recognizable subsets are the same in
both interpretations, since their definition is given in terms of the
same algebraic structure, that of the free commutative monoid over
$A$, but recognizability is equivalent to $\CMSO$-definability in one
interpretation, and to $\MSO$-definability in the other.

Say that a map $\phi\colon S\rightarrow T$ between sets of relational
structures is a $\MS$-transduction if there exist $\MSO$-formulas (in the
language of the relational structures in $S$) that express each $s\phi$
(its domain and its relations) as a subset of a direct product of a
fixed number of copies of $s$ (see Courcelle \cite{BCHdbk} for a
precise definition). For instance, if $A_{0}$ is the subset
of constants in an alphabet $A$, the word in $A_{0}^*$ formed by the
leaves of an $A$-labeled tree $t$ can be easily described by
$\MSO$-formulas inside the set of nodes of $t$.

Now consider a set of relational structures $M$, equipped with an
algebraic structure with signature $\Sigma$. A simple example is given
by $M = A^*$, the set of words on alphabet $A$, seen as a monoid: the
signature $\Sigma$ consists of a binary operation (interpreted in $A^*$
as concatenation) and of $|A|$ constant symbols (interpreted in $A^*$
as the letters of $A$). The valuation morphism $\val$ maps every
$\Sigma$-term (a $\Sigma$-labeled tree) to its interpretation in $M$.
The following result is due to Courcelle \cite{BCI,BCHdbk}.

\begin{theorem}\label{BCI}
    If the valuation morphism is surjective and is an
    $\MS$-trans\-duction, then every $\CMSO$-definable subset of $M$ is
    recognizable.
\end{theorem}

The mechanism of the proof is worth sketching: let $L\subseteq M$ be 
$\CMSO$-definable. The inverse image of a $\CMSO$-definable set by an
$\MS$-transduction is $\CMSO$-definable, so $\val\inv(L)$ is
$\CMSO$-definable in $T_{\Sigma}$. But in the set of $\Sigma$-labeled 
trees, $\CMSO$-definability is equivalent to $\MSO$-definability, and
hence to recognizability. And it is easy to show that if
$\val\inv(L)$ is recognizable, then so is $L$.

Examining this sketch of proof also sheds light on the decidability of
$\CMSO$-defined sets, and on the complexity of such a decision problem.
Suppose we have a parsing algorithm, which maps a given relational
structure $x\in M$ to a $\Sigma$-term $\parse(x)$ describing it. Let
$L\subseteq M$ be described by a $\CMSO$-formula $\phi$ and let $x\in
M$: we want to decide whether $x\in L$. An $\MSO$-formula $\psi$
describing $\val\inv(L)$ can be computed from $\phi$ and the formulas
describing the $\MS$-transduction $\val$. The problem then reduces to
deciding whether $\parse(x)$ satisfies $\psi$, and by
Theorem~\ref{TWD}, this can be solved (efficiently) by running
$\parse(x)$ through a bottom-up tree automaton.

The converse of Theorem~\ref{BCI} does not always hold: there are
situations, in particular in the discussion of languages of graphs,
where some recognizable sets are not $\CMSO$-definable. However, the
two notions are known to be equivalent in important cases: we have
already seen it for words or trees; other interesting situations are
discussed in Sect.~\ref{sec poset} and~\ref{sec graph}. It is
interesting to note that a common feature of those situations where the
notions of definability and recognizability coincide, is that we are
able to describe a parsing function $\parse$ as an $\MS$-transductions,
and this is possible only because some finite generation condition is
assumed to hold (which cannot be assumed for the class of all finite
graphs).

For some of the specific relational structures discussed in the sequel,
there is a notion of automaton that matches the definition of
recognizability -- but in many other situations, especially when
dealing with graphs or posets, no such notion is known. In those cases,
the algebraic approach is really the only tool we have to characterize
logical definability, and to hope to bring about decision algorithms.

\section{Recognizable Sets of Discrete Structures}\label{successful examples}

For the discrete structures discussed in this section, fruitful
algebraic structures have been introduced in the literature. The first
measure of the interest of such algebraic structures, is whether the
corresponding notion of recognizability matches some natural notion of
logical definability, or some natural notion of recognizability by
automata. A second measure of interest is whether the algebraic theory
thus introduced allows us to characterize -- and if possible decide --
significant classes of recognizable sets. Typically, deciding
$\FO$-definability is a key problem, but other classes may arise
naturally depending on the type of discrete structures we consider.

\subsection{Infinite Words}\label{sec infinite words}

We start with infinite words because it is an area where the theory has
been developed for a long time (B\"uchi's theorem goes back to the
early 1960s), and has a strong algebraic flavor. Here we are talking of
\textit{one-way infinite words}, or \textit{$\omega$-words}, that is,
$A$-labeled infinite chains, or elements of $A^\N$. For a detailed
presentation of the results surveyed in this section, we refer the
readers to Perrin and Pin's book \cite{PPbook} and to the survey
papers \cite{PPNATO,Pin1996}.

The notions of \textit{B\"uchi} and (deterministic) \textit{Muller} or
\textit{Rabin automata} were evolved in the 1960s, and they were proved
to have the same expressive power as $\MSO$-formulas (on $A$-labeled
infinite chains), and as \textit{$\omega$-rational expressions}. The
latter describe every $\MSO$-definable language of $\omega$-words as
finite unions of products of the form $KL^\omega$, where $K,L$ are
recognizable languages of finite words and the $\omega$-power denotes
infinite iteration. In particular, this indicates that the sets of
$\omega$-words that can be accessed by $\MSO$ or automata-theoretic
specifications are in a sense ultimately infinite iterations of a
recognizable set of finite words.

An algebraic approach to $\omega$-rational languages took longer to
evolve. Early work of Arnold, P\'ecuchet, Perrin emphasized the
necessary interplay of relations on $A^\N$ (concerning infinite words)
and ordinary monoid congruences on $A^*$ (concerning finite words). It
also emphasized that nothing much could be expected from the monoid
structure of $A^\N$, in which every product is equal to its first
factor. Eventually, it was recognized that finite and infinite words
cannot be considered separately, but they form a two-sorted algebra, as
explained in Sect.~\ref{multisorted}. The definition of the binary
concatenation product does not pose any problem, but must be split in
one operation of type $\textsf{finite}^2\rightarrow\textsf{finite}$ and
one operation of type
$\textsf{finite}\times\textsf{infinite}\rightarrow\textsf{infinite}$.
But the generation of infinite words from finite one can be envisaged
in two fashions: we can consider an $\omega$-ary product, of type
$\textsf{finite}^\omega\rightarrow\textsf{infinite}$, or the unary
$\omega$-power operation, of type
$\textsf{finite}\rightarrow\textsf{infinite}$. The first choice is
termed an \textit{$\omega$-semigroup} (Perrin, Pin), and $A^\infty = A^+\cup
A^\N$ is (freely) generated by $A$ as an $\omega$-semigroup. The second
choice is termed a \textit{Wilke algebra} (Wilke), and the sub-Wilke algebra of
$A^\infty$ generated by $A$ consists in the finite and ultimately
periodic $\omega$-words only. A Ramsey theorem shows however that on a
finite set, a Wilke algebra structure can be canonically extended to an
$\omega$-semigroup structure, so that the consideration of these two
algebraic structures yields the same class of recognizable languages.

The robustness of this algebraic approach to recognizable subsets of
$A^\infty$ (and not $A^\N$!) is such that an Eilenberg-style theory of
varieties was developped (see Sect.~\ref{sec words}), and that a good
number of combinatorially or logically interesting classes of
recognizable sets have been characterized algebraically (Perrin and Pin
\cite{PPbook}, Carton \cite{Carton00}).

From the algorithmic point of view, note that passing from a B\"uchi
automaton to a deterministic Muller or Rabin automaton (say, for the
purpose of complementation) is notoriously difficult, see Safra's
exponential time algorithm, but no significantly better algorithm is
possible \cite{ThomasHdbk}.

Elegant results generalize this discussion to transfinite words, that
is, $A$-labeled ordinals longer than $\omega$, see the work of Bedon,
Bruy\`ere, Carton, Choueka \cite{Choueka,BedonJCSS,BedonIJAC,BruCar02}.

For infinite trees, we know models of automata that are equivalent to
$\MSO$-decidability (Rabin, see \cite{ThomasHdbk}), but the extension
of the algebraic ideas sketched above remains to be done. The
finiteness results implied by Ramsey's theory seem much harder to
obtain for trees.

\subsection{Poset-Related Models}\label{sec poset}

A pomset (partially ordered multiset) is an $A$-labeled poset. The
first example, of course, is that of words, which are $A$-labeled
chains. Other examples were considered, and first of all the case of 
traces.

\subsubsection{Traces.}
There the alphabet is equipped with a structure -- which can be viewed
as an independence relation, or a dependence relation, or a distributed
structure. A \textit{trace} can then be viewed in several fashions: as
an equivalence class of words in the free monoid $A^*$, in the
congruence induced by the commutation of independent letters (so traces
form a monoid); or as a so-called \textit{dependence graph}, that is,
an $A$-labeled poset where the order is constrained by the distributed
structure of the alphabet, see Diekert and Rozenberg's book
\cite{DR95}. The latter is the more significant model, from the point
of view of the original motivation of traces as a model of distributed
computation.

The power of $\MSO$-definability -- interpreted on the dependence graph
model -- was proved to be equivalent to the power of \textit{Zielonka's
automata} (a model of automata which which incorporates information on
the distributed structure of the alphabet), and to algebraic
recognizability in the trace monoid \cite{DR95}.

Note that, as discussed in Sect.~\ref{def rec}, in the particular
case where the letters are independent from one another, the trace
monoid is the free commutative monoid. When elements of this monoid are
represented by trace dependence graphs, where for each letter $a\in A$,
the set of $A$-labeled elements is a chain, then antichains have
bounded cardinality (that of $A$), and a linearization of the poset can
be defined by a $\MSO$-formula, so $\MSO$-definability is equivalent to
$\CMSO$-definability. When the elements of the same monoid are
represented by finite discrete $A$-labeled graphs, without any edges,
then $\MSO$-definability is strictly weaker than $\CMSO$-definability. In
both cases however, recognizability is equivalent to
$\CMSO$-definability.

Good results are also known for $\FO$-definable trace languages: they
are characterized by star-free rational expressions, and by the
aperiodicity of their syntactic monoid (Guaiana, Restivo, Salemi
\cite{GRS}), and important temporal logics with the same expressive
power have been developed (see Thiagarajan and Walukiewicz
\cite{ThWa97} and Diekert and Gastin \cite{DG04}).

There is a large body of literature on recognizable trace languages,
and the results summarized above point to a rather well understood
situation. Some questions however are not solved in a completely
satisfactory fashion. For instance, the question of rational
expressions for trace languages remains unclear (see the star problem):
the difficulty comes from the fact that the star of a recognizable
trace language may not be recognizable; the notion of concurrent star,
which takes care of that obstacle, retains an \textit{ad hoc} flavor
\cite{DR95}. Similarly, with the remarkable exception of
$\FO$-definable trace languages, the task of identifying,
characterizing and deciding interesting subclasses of recognizable
languages has eluded efforts.

One can argue that this is due to the loss of information that occurs
if we consider the set of traces as a monoid -- which we must do if the
algebraic structure on the set of traces is that of a monoid: in the
resulting definition of recognizability, a set of traces is
recognizable if and only if its set of linearizations (in $A^*$) is
recognizable. From an algebraic point of view, this puts the emphasis
on commutation, but two traces may commute because they are
independent, or because they are powers of a third one, in which case
they are deeply dependent. From a more algorithmic point of view, what
is done there is to reduce the study of a trace language to the study
of the language of all its linearizations.

On the other hand, Zielonka's automata succeed in taking into account
the distributed structure of the computation model, and are
well-adapted to traces. Since they match monoid recognizability all the
same, this points to the following problem: to find an alternative
algebraic structure on the set of traces, which does not change the
family of recognizable sets, yet better accounts for the distributed
nature of that model, and hence (hopefully) naturally connects with
Zielonka's automata (\textit{i.e.}, provides an algebraic proof of
Zielonka's theorem) and allows the identification and characterization
of structurally significant subclasses of recognizable trace languages.

Infinite traces exhibit interesting properties, from the point of view
of auto\-mata-recognizability and logical definability, see
\cite{DR95,EbingerMuscholl}.

\subsubsection{Message Sequence Charts and Communication Diagrams.}

\textit{Message sequence charts} (\textit{MSC}s) form a specification
language for the design of communication protocols, that has attracted
a lot of attention in the past few years. They can also be considered
as specifications of particular pomsets, that are disjoint unions of
$k$ chains. An abstraction of this model is given by \textit{Lamport
Diagrams} (LDs) and by \textit{Layered Lamport Diagrams} (LLDs), which
are LDs subject to a boundedness condition.

Henriksen, Kumar, Mukund, Sohoni, Thiagarajan \cite{HMKT1,HMKT2,MKS}
considered the class of \textit{bounded} finite MSC languages, defined
by so-called \textit{bounded} (Alur, Yannakakis \cite{AY99}) or
\textit{locally synchronised} (Muscholl, Peled \cite{MP-MFCS99})
MSC-graphs. For bounded MSC languages, $\MSO$-definability is
equivalent to rationality of the language of all linearizations, and to
recognizability by deterministic (resp. non-deterministic)
\textit{message-passing automata}. Kuske extended these results to
$\FO$-definable MSC languages, and to infinite bounded MSCs
\cite{KuskeMSC}.

The restriction to classes of posets with a rational language of
linearizations is rather severe, but little work so far has discussed
definability or recognizability outside this hypothesis. Meenakshi and
Ramanujam \cite{Meenakshi} and Peled \cite{Peled2000} investigated
decidable logics for MSCs and LLDs, that are structural, \textit{i.e.},
not defined on the language of linearizations. There does not seem yet
to exist an algebraic approach of (a subclass of LDs) that would match
the power of $\MSO$-definability.

\subsubsection{Series-Parallel Pomsets.}
Sets of series-parallel pomsets (or \textit{$sp$-languages}) were
investigated by Lodaya, Weil, Kuske \cite{KuskeSP,LW-TCS}. A poset is
\textit{series-parallel} if it can be obtained from singletons by using
the operations of sequential and parallel product. There is a
combinatorial characterization of these posets ($N$-free posets
\cite{Grabowski,VTL81}), but the definition above naturally leads to
the consideration of the so-called \textit{series-parallel algebras}
\cite{LW-TCS}, that is, sets equipped with two binary associative
operations, one of which is commutative. Kuske showed that an
$sp$-language is recognizable if and only if it is $\CMSO$-definable
\cite{KuskeSP}. Lodaya and Weil introduced a model of branching
automata and a notion of rational expressions, which they proved had
the same expressive power \cite{LW-TCS}. However these automata accept
not only the recognizable $sp$-languages, but also some
non-recognizable ones.

The \textit{bounded-width} condition is a natural constraint on
$sp$-languages: a set $L$ of series-parallel pomsets has bounded-width
if there is a uniform upper bound on the cardinality of an anti-chain
in the element so $L$. Results of Kuske, Lodaya and Weil
\cite{KuskeSP,LW-TCS} show that when we consider only bounded-width
$sp$-languages, then recognizability is equivalent to
automata-recognizability, to $\MSO$-decidability, and to expressibility
by a so-called \textit{series-rational expression}. Under the
boun\-ded-width hypothesis, $\FO$-definable $sp$-languages are
characterized by a notion of star-free rational expressions, and by an
algebraic condition on the syntactic $sp$-algebra which is analogous to
the aperiodicity of monoids \cite{KuskeSP}.

\subsubsection{Texts and $n$-Pomsets.}
An \textit{$A$-labeled text} is a finite $A$-labeled set, equipped with
2 linear orders. Texts form a particular class of the
\textit{2-structures} studied by Ehrenfeucht, Engelfriet, Harju,
Proskurowski and Rozenberg \cite{ER90,ER92,EHPR96}. Hoogeboom and ten
Pas introduced an algebraic structure on the set of all texts
\cite{HtP1}. This algebra has an infinite signature, but within any
finitely generated sub-algebra (generated by $A$ and any finite subset
of the signature, the hypothesis of \textit{bounded primitivity} in
\cite{HtP1,HtP2}), recognizability is equivalent to
$\MSO$-definability.

The class of texts generated by the alphabet and the two arity 2
operations on texts (\textit{alternating texts}) is of particular
interest, as we now discuss.

A pair of linear orders $(\le_{1},\le_{2})$ on a finite set specifies
and can be specified by a pair of partial orders $(\sqsubseteq_{1},
\sqsubseteq_{2})$ such that every pair of distinct elements is
comparable in exactly one of these partial orders (this defines a
\textit{2-poset}): it suffices to take ${\le_{1}} = {\sqsubseteq_{1}}
\cup{\sqsubseteq_{2}}$ and ${\le_{2}} = {\sqsubseteq_{1}}
\cup{\sqsupseteq_{2}}$; and conversely ${\sqsubseteq_{1}} = {\le_{1}}
\cap{\le_{2}}$ and ${\sqsubseteq_{2}} = {\le_{1}} \cap{\ge_{2}}$. Since
the translation between texts and 2-posets is described by
quantifier-free formulas, $\MSO$-definability is preserved under this
translation. On 2-posets, one can consider two natural operations: one
behaves like a sequential product on $\sqsubseteq_{1}$ and a parallel
product on $\sqsubseteq_{2}$; and the other is defined dually,
exchanging the roles of the two partial orders. Let $SPB(A)$ be the
algebra of 2-posets generated by $A$ and these two operations. \'Esik
and N\'emeth observed that these two operations on 2-posets translate
to the two arity 2 operations of the text algebra; moreover, they
introduced a simple model of automata for subsets of $SPB(A)$, whose
power is equivalent to recognizability and to $\MSO$-definability
\cite{EsikNemethK-dim}.

\'Esik and N\'emeth's automata can also be defined for $n$-posets,
where they are also equivalent to recognizability and to
$\MSO$-definability.

\subsubsection{Pomsets in General.}
There does not seem to be a natural model of automaton that makes sense
on all pomsets. However, since pomsets can be represented by
$A$-labeled directed acyclic graphs (\textit{dags}), they are directly
concerned by the discussion in the next section. In particular, and
getting ahead of ourselves, let us observe that the subsignature of the
modular signature consisting of the operations defined by graphs that
are posets (resp. dags), generates the class of all finite posets
(resp. dags) - and the results on $\CMSO$-definability discussed in
Sect.~\ref{sec graphs} therefore apply to pomset languages.

\subsection{Graphs and Relational Structures}\label{sec graph}\label{sec
graphs}

Graphs (edge- or vertex-labeled, colored, with designated vertices,
etc), and beyond them, relational structures (\textit{i.e.},
hypergraphs) are the next step, and they occur indeed in many modeling
problems. The notion of logical definability is rather straightforward,
although it may depend on the logical structure we consider on graphs
(whether a graph is a set of vertices with a binary edge predicate, or
two sets of vertices and edges with incidence predicates). From the
algebraic point of view, there is no prominent choice for a signature
to describe graphs. However, three signatures emerge from the
literature. One of them, the modular signature, arises from the theory
of modular decomposition of graphs, the other two (the $HR$- and the
$VR$-signature) arise from the theory of graph grammars. We will also
consider a fourth signature, on the wider class of relational
structures. We will see that under suitable finiteness conditions, the
resulting notions of recognizability are equivalent.

After briefly describing these signatures, and comparing the notions of
recognizability which they induce, we rapidly survey known definability
results. We conclude with the discussion of a couple of situations
where automata-theoretic models have been introduced.

\subsubsection{The Modular Signature.}
 
A concrete graph $H$ with vertices $\{1,\ldots,n\}$ and edge set
$E_{H}$, induces an $n$-ary operation on graphs as follows: the vertex
set of the graph $H\langle G_{1},\ldots,G_{n}\rangle$ is the disjoint
union of the vertex sets of the $G_{i}$, it contains all the edges of
the $G_{i}$, and for each edge $(i,j)\in E_{H}$, it also has all the
edges from a vertex of $G_{i}$ to a vertex of $G_{j}$. A graph is said
to be \textit{prime} if it cannot be decomposed non-trivially by such
an operation. The \textit{modular signature} $\calF_{\infty}$ consists
of the set of all prime graphs, or rather, of one representative of
each isomorphism class of prime graphs. It is an infinite signature. In
particular, $\calF_{\infty}$ contains a finite number of operations of
each arity. The operations of arity 2 are the parallel product, the
sequential product and the clique product: they are defined by the
graphs with 2 vertices and no edge, 1 edge and 2 edges, respectively
\cite{BCX,Weil04}.

The theory of modular decomposition of graphs
\cite{MS99,Mohring-Radermacher} shows that the class of all finite
graphs is generated by the singleton graph and $\calF_{\infty}$, and
describes the relations between the operations in $\calF_{\infty}$.
Finite $A$-labeled graphs are generated by $A$ and $\calF_{\infty}$. If
$\calF$ is a finite subset of $\calF_{\infty}$, the algebra generated
by $A$ and $\calF$ is called the class of $A$-labeled $\calF$-graphs.

For instance, if $\calF$ consists of the sequential product, the
$\calF$-graphs are the finite words. If $\calF$ consists of the
parallel (resp. clique) product, they are the discrete graphs (resp.
cliques). If $\calF$ consists of the parallel and the sequential
products, we get the series-parallel posets (see Sect. \ref{sec
poset}), and if it consists of the parallel and clique products, we
get the cographs (see below).

\subsubsection{The Signature $HR$.}

Here, graphs are considered as sets equipped with a binary edge
predicate, and a finite number of constants (\textit{i.e.}, designated
vertices), called \textit{sources}. Each finite set of source names
defines a sort in the $HR$-algebra $\GS$ of \textit{graphs with
sources}. The operations in the algebra are the disjoint union of
graphs with disjoint sets of source names, the renaming of sources,
forgetting sources, and the fusion of two sources \cite{BCHdbk}. A
number of variants can be considered, which do not affect the class of
$HR$-recognizable subsets, see Courcelle and Weil \cite{BCPW}: the
disjoint union can be replaced with parallel composition (source name
sets need not be disjoint, and sources with the same name get
identified), the sources may be assumed to be pairwise distinct
(\textit{source separated graphs}), the source renaming operations can
be dropped, or the source forgetting operations, etc.

The signature $HR$ emerged from the literature on graph grammars, and
the acronym $HR$ stands for \textit{Hyperedge Replacement}. More
precisely, the equational sets of graphs, relative to the signature
$HR$, are known to enjoy good closure properties, and can be elegantly
characterized in terms of recognizable tree languages and
$\MS$-transductions where both vertex and edge sets can be quantified
(see Courcelle \cite{BCHdbk}).

\subsubsection{The Signature $VR$.}

Now graphs are considered as sets equipped with a binary edge
predicate, and a finite number of unary predicates (\textit{i.e.},
colors on the set of vertices), called \textit{ports}. Each finite set
of port names defines a sort in the $VR$-algebra $\GP$ of
\textit{graphs with ports}. The operations in the algebra are the
disjoint union, the edge adding operation (adding an edge from each
$p$-port to each $q$-port for designated port names $p,q$), and the
renaming and forgetting of port names. Again, the class of
$VR$-recognizable subsets is not affected by variants such as the
consideration of graphs where ports must cover the vertex set, or must
partition it. It also coincides with $\NLC$-recognizable
graphs, see Courcelle and Weil \cite{BCPW}.

The signature $VR$ (standing for \textit{Vertex Replacement}) also
emerged from the literature on graph grammars, and the equational sets
of graphs, relative to the signature $VR$, enjoy good closure
properties, and are characterized in terms of recognizable tree
languages and $\MS$-transductions where only vertex sets can be
quantified (see Courcelle \cite{BCHdbk}).

\subsubsection{The Signature $\calS$ on Relational Structures with Sources.}

Subsuming the algebras of graphs with sources and with ports, we can
consider the class of \textit{relational structures with sources}
$\StS$. These are sets equipped with a finite relational structures,
and a finite number of constants (\textit{sources}). Each pair
consisting of a relational signature and a set of source names defines
a sort, and the operations in the signature $\calS$ are the disjoint
union between sorts with disjoint sets of source names, and all the
unary operations that can be defined on a given sort using
quantifier-free formulas, see \cite{BCPW}. The operations in the
signatures $VR$ and $HR$ are particular examples of such
\textit{quantifier-free definable operations}. The notion of
$\calS$-recognizability is not affected if we consider parallel
composition instead of disjoint union (as for the signature $HR$), nor
if we consider only structures where sources are separated \cite{BCPW}.

\subsubsection{Comparing the Notions of Recognizability.}

Combining a number of results of Courcelle and Weil \cite{BCPW}, we
find that a set of graphs is $VR$-recognizable if and only if it is
$\calS$-recognizable. Moreover, a $VR$-recognizable set of graphs is
$\calF_{\infty}$-recognizable, and it is also $HR$-recognizable (the
implication for $HR$- and $VR$-equational sets goes in the other
direction). Finally, $\calF_{\infty}$-, $HR$- and $VR$-recognizability
(resp. equationality) are equivalent under certain boundedness
conditions.

In particular if we consider a set $L$ of graphs without $\vec K_{n,n}$
for some $n$ ($\vec K_{n,n}$ is the complete bipartite directed graph
with $n+n$ vertices), then $L$ is $VR$-recognizable if and only if it
is $HR$-recognizable \cite{BCPW}. This sufficient condition is implied
by the following boundedness properties (in increasingly general
order): the graphs in $L$ have uniformly bounded degree, they have
bounded tree-width, they are sparse. The notion of bounded tree-width
can be seen as a finite generation property, relative to the signature
$HR$ \cite{BCHdbk}.

If $\calF$ is a finite subset of the modular signature $\calF_{\infty}$
and $L$ is a set of $\calF$-graphs, then $L$ is $\calF$-recognizable if
and only if it is $\calF_{\infty}$-recognizable (resp.
$VR$-recognizable, $\calS$-recognizable) \cite{BCX,BCPW}.

\subsubsection{Monadic Second-Order Definability.}

From the logical definability point of view, graphs can be seen as sets
(of vertices) equipped with an edge predicate, or as pairs of sets (of
vertices and edges respectively) with incidence predicates. Let us
denote by $\CMSO[\edge]$ the $\CMSO$ emerging from the first point of
view, and by $\CMSO[\inc]$ the second one. It is easily verified that
$\CMSO[\edge]$-definable sets of graphs are also
$\CMSO[\inc]$-definable. Moreover $\CMSO[\edge]$-definability implies
$VR$-recognizability, and $\CMSO[\inc]$-definability implies
$HR$-recognizability, see Courcelle \cite{BCHdbk}.

Lapoire showed that if $L$ is a set of graphs with bounded tree-width,
then $\CMSO[\inc]$-definability is equivalent to $HR$-recognizability
\cite{Lapoire}. Moreover, if $L$ is uniformly sparse, then
$\CMSO[\edge]$- and $\CMSO[\inc]$-defin\-ability are equivalent
(Courcelle \cite{BCXIV}). In view of the equivalence result between
$HR$- and $VR$-recog\-nizability mentioned above, it would be interesting
to find out whether both definabilities are also equivalent if $L$ is
without $\vec K_{n,n}$ for some $n$.

Returning to the modular signature, $\CMSO[\edge]$-definability implies
$\calF_{\infty}$-recog\-nizability, by general reasons (see
Sect.~\ref{def rec}). Weil showed that the converse holds for sets of
$\calF$-graphs, provided $\calF$ is finite and the operations of
$\calF$ enjoy a limited amount of commutativity (\textit{weakly rigid
signatur}e, see \cite{Weil04} for details). This assumption is rather
general, and covers in particular all the cases where $\calF$-graphs
are dags or posets, and notably the case of $sp$-languages.

A typical example of a subsignature of $\calF_{\infty}$ which is not
weakly rigid, consists of the parallel and the clique products, two
binary commutative, associative operations which generate the cographs.
\textit{Cographs} form a class of undirected graphs, closely related
with comparability graphs (Corneil, Lerchs, Stewart \cite{Corneil81})
and can be characterized as follows: an undirected graph is a cograph
if and only if it does not contain $P_{4}$ ($P_{4}$ has vertex set
$\{1,\ldots,5\}$ and edges between $i$ and $i+1$, $1\le i\le 4$). The
arguments that show that $\calF$-recognizability is equivalent to
$\CMSO[\edge]$-definability when $\calF$ is weakly rigid, fail for
cographs. Courcelle \cite{BCX} asks whether $\CMSO$-definability is
strictly weaker than $\calF$-recognizability for a general finite
subsignature $\calF \subseteq \calF_{\infty}$: the first place to look
for a counter-example seems to be cographs.

\subsubsection{Series-$\Sigma$ Algebras.}

In their investigation of $sp$-languages, Lodaya and Weil introduced
series-$\Sigma$-algebras and their subsets ($s\Sigma$-languages): here
$\Sigma$ is a ranked alphabet (as in the study of tree languages,
Sect.~\ref{sec trees}) and $\bullet$ is a binary associative operation
not in $\Sigma$. The \textit{$s\Sigma$-terms}, that is, the elements of
the algebra freely generated by $A$, $\Sigma$ and $\bullet$ (called the
\textit{free series-$\Sigma$-algebra over $A$}), can be viewed as
finite sequences of $\Sigma$-trees, where each child of the root is in
fact a smaller $s\Sigma$-term. Lodaya and Weil introduced a model of
automata and a notion of rational expressions, both of which are
equivalent to recognizability \cite{LW-IC} -- a result which
generalizes the characterization of recognizability by finite automata
for both words and trees. They showed that their result could be
adapted if some of the operations in $\Sigma$ were assumed to be
commutative, but not if some amount of associativity was introduced
(e.g. $sp$-languages, cographs). The logical dimension of
$s\Sigma$-languages was not developped.

\subsubsection{Automata for Graph Languages.}

We have seen some automata, designed for specific situations (finite
and infinite words, traces, series-parallel pomsets, MSC languages,
$n$-pomsets), see Sect.~\ref{sec poset}. As discussed there, these
automata models match the expressiveness of recognizability and
$\MSO$-defin\-ability, sometimes under additional boundedness
hypothesis.

For general graph languages, Thomas introduced the notion of
\textit{graph acceptor} \cite{ThomasTAPSOFT97,ThomasHdbk97},
generalizing the \textit{tiling systems} introduced earlier by
Giammarresi and Restivo \cite{GR97} for \textit{pictures} ($A$-labeled
rectangular grids). Recognizability by a graph acceptor was shown to be
equivalent to $E\MSO$-definability, where $E\MSO$ is the extension of
$\FO$ by existential quantification of monadic second order variables.

\subsection{Revisiting Trees}

As mentioned in Sect.~\ref{sec trees}, the problem of deciding whether
a given recognizable set of $\Sigma$-trees is $\FO$-definable is still
open, and various attempts to use the structure of $\Sigma$-algebras
described in Sect.~\ref{sec trees} to solve it in the spirit of
Sch\"utzenberger's theorem (Theorem~\ref{SMcNK}), have failed
\cite{Heuter,Potthoff1,Potthoff95}. Recently, \'Esik and Weil
introduced a new algebraic framework to investigate this particular
problem on tree languages \cite{EsikWeil}. The point was to enrich the
algebraic framework, without modifying the notion of a recognizable
subset, but introducing additional algebraic structure.

\'Esik and Weil's algebras, called \textit{preclones}, are multi-sorted
algebras with one sort for each integer $n$, and $\Sigma$-trees form
the $0$-sort of the free $\Sigma$-generated preclone. As indicated, a
set of $\Sigma$-trees is preclone-recognizable if and only if it is
recognizable with respect to $\Sigma$-algebras; moreover, if $L$ is a
recognizable $\Sigma$-tree language, its syntactic preclone (which is
finitary but not finite due to the infinite number of sorts) admits a
finite presentation, encoded in the minimal bottom-up automaton of $L$.
This is naturally important if we want to use syntactic preclones in
algorithms.

The main result of \cite{EsikWeil} states that $L$ is $\FO$-definable
if and only if its syntactic preclone lies in the least pseudovariety
of preclones closed under \textit{two-sided wreath products}, and
containing a certain very simple 1-generated preclone. The two-sided
wreath product is a generalization of the operation of the same name on
monoids (Rhodes, Tilson, see \cite{Straubing}), and this result
generalizes Sch\"utzenberger's theorem on finite words. It is the first
algebraic characterization of $\FO$-definable tree languages, but
unfortunately, it is not clear at this point whether this
characterization can be used to derive a decision algorithm.

The approach in \cite{EsikWeil} also applies to $\FO{+}\MOD$-definable
tree languages, and other similarly defined languages.

\subsection{Timed Models}
 
\textit{Timed automata} appeared in the 1990s, to represent the
behavior of finite state systems subjected to explicit time constraints
(Alur, Dill \cite{AD94}). While they are already widely used, the
foundations of the corresponding theory are still under development.
There are several variants of these automata, such as event-clock
automata, and of the models of timed computations (timed words, clock
words, etc). There have also been several attempts to develop
appropriate notions of rational expressions, that would be equivalent
to the expressive power of timed automata, see Henzinger, Raskin,
Schobbens \cite{HRS}, Asarin, Maler, Caspi \cite{AMC}, Dima
\cite{Dima01}, Maler, Pnueli \cite{MP} among others. At the same
time, timed automata and timed languages may exhibit paradoxical
behaviors, due to the continuous nature of time, so the central ideas
and techniques from the classical theory cannot simply be enriched with
timed constraints to account for the behavior of timed automata.

The development of an algebraic apparatus and of a logical formalism is
also still in its infancy. One should mention however the recent work
of Maler and Pnueli \cite{MP}, and the results of Francez and Kaminski
\cite{FK} and Bouyer, Petit and Th\'erien \cite{BPT} on generalizations
of timed languages and automata, to \textit{automata on infinite
alphabets} and to \textit{data languages}, respectively. In both cases,
an interesting notion of algebraic recognizability is introduced, that
is at least as powerful as timed automata. languages. Moreover, several
logics have been introduced (see for instance Demri, D'Souza
\cite{DDS}), but none is completely satisfactory with respect to the
motivation of formulating and solving the controller synthesis problem,
and none is connected in a robust way to an algebraic approach of
recognizability.

The  development of a complete theory of timed systems, incorporating
aut\-omata-theoretic, algebraic and logical aspects, appears to be one 
of the more difficult challenges of the moment.

%


\end{document}